\documentclass[dvips]{elsart}

\usepackage{epsfig,graphicx,amstext,psfrag,array}

\begin{document}

\journal{Nucl. Phys. A}

\begin{frontmatter}

\title{Production cross-sections and momentum distributions of fragments from 
neutron-deficient $^{36}$Ar at 1.05 A.GeV}

   \author[usc]{M.~Caama{\~n}o},
   \author[usc,gsi]{D.~Cortina-Gil},
   \author[gsi]{K~.S{\"u}mmerer},
   \author[usc]{J.~Benlliure},
   \author[usc]{E.~Casarejos},
   \author[gsi]{H.~Geissel},
   \author[gsi]{G.~M{\"u}nzenberg},
   \author[usc]{J.~Pereira}
   \address[usc]{Universidad de Santiago de Compostela,
     15706 Santiago de Compostela, Spain}
   \address[gsi]{Gesellschaft f{\"u}r Schwerionenforschung,
     Planckstr.1, 64291 Darmstadt, Germany}


\begin{abstract}
We have measured production cross sections and longitudinal momentum distributions of fragments from neutron-deficient $^{36}$Ar at 1.05 {\it A}GeV. The production cross-sections show excellent agreement with the predictions of the semiempirical formula {\sc EPAX}. We have compared these results, involving extremely neutron deficient nuclei, with model calculations to extract information about the response of these models close to the driplines. The longitudinal momentum distributions have also been extracted and are compared with the Goldhaber and Morrissey systematics. 
\end{abstract}

\end{frontmatter}

\section{Introduction}
Projectile fragmentation at high  energies ($\approx 1~ A$ GeV)
has proven to be a powerful tool to produce clean secondary beams of exotic nuclei
that can be used for a wide variety of nuclear-structure, nuclear-reaction, or nuclear-astrophysics
studies.
To make full use of this potential of projectile fragmentation, the reaction mechanism has to be understood in
detail in order to predict precisely the properties of the secondary beams. 
Pioneering experiments were performed at the LBL Bevalac accelerator~\cite{viyogi,ca48}.
Later, more detailed studies have been undertaken at the SIS/FRS exotic beam facility at GSI, Darmstadt
\cite{blank,weber,joerg,dejong}.
Comprehensive datasets have been obtained with respect to formation cross sections of exotic fragments
as well as their kinematic properties (mainly centroids and widths of longitudinal momentum distributions).
Both types of observables could be fitted by empirical parametrizations:
the former by the widely used {\sc EPAX} parametrization~\cite{epax}, the latter by the 
Morrissey systematics~\cite{morrissey}. These parametrizations have turned out to be useful within
simulation programs of projectile fragment separators (e.g. {\sc MOCADI}~\cite{mocadi} or {\sc LISE}~\cite{lise}).
More insight into the underlying physical processes was provided by comparison to physical
models like abrasion-ablation ({\sc ABRABLA}~\cite{abra,Jun98}) or intranuclear-cascade models ({\sc ISABEL}
~\cite{Isabel}).

Since most projectile-fragmentation experiments aim at producing exotic nuclei near the limits of
the presently known regions, and since the fragment neutron- or proton-excess is correlated with the
neutron- or proton-excess of the projectile~\cite{epax}, fragmentation of stable projectiles
outside the valley of $\beta$-stability has received most interest. Quite a few studies of
neutron-rich projectiles like $^{48}$Ca, $^{86}$Kr, and $^{136}$Xe have been reported in the
literature~\cite{ca48,weber,xe136}. On the proton-rich side, fragmentation cross sections for
$^{58}$Ni, $^{112}$Sn, and $^{124}$Xe have been published~\cite{blank,stolz,xe124}.
It was found that the empirical {\sc EPAX} parametrization could fit the measured cross sections 
with good accuracy down to sub-nanobarn levels~\cite{epax}.

The present paper aims at complementing the data base for the fragmentation of proton-rich
projectiles by studying $^{36}$Ar fragmentation. The isotope $^{36}$Ar is important for 
producing light proton-rich nuclei at the proton drip line; their properties can be compared to those from $^{40}$Ar fragmentation~\cite{viyogi}. In addition to the formation
cross sections (down to a level of $\mu$barn), widths of longitudinal momentum distributions
could be measured; both will be compared in the following to the respective systematics as well
as to physical model predictions.

\section{Experimental procedures}
The results reported in the present paper were obtained as a by-product of a secondary-reaction
experiment aimed at measuring total interaction cross sections of Ar and Cl isotopes
\cite{Oza02} as well as measuring longitudinal momentum distributions of knock-out products
in coincidence with $\gamma$-rays~\cite{Lola03}.
This is the reason for certain experimental limitations of the present experiment that will be discussed
below.
The experiment was undertaken at the SIS/FRS facility at GSI in Darmstadt, Germany~\cite{Gei92}.
SIS delivered a $^{36}$Ar beam with 1050 $A$ MeV in spills of 8 $s$ duration with a
repetition rate of 1/16 $s^{-1}$.
Typical beam intensities varied between $10^{8}$ and $10^{9}$ ions per spill, depending on the
cross section to be measured.
The primary-beam intensities were measured with a secondary-electron transmission monitor
(SEETRAM)~\cite{Jur02} that was calibrated with respect to a fast scintillator detector.
Fragments were produced in a thick (1625 mg/cm$^2$) $^9$Be target located at the entrance
of the FRS.    

\begin{figure}
  \begin{center}
   \epsfig{file=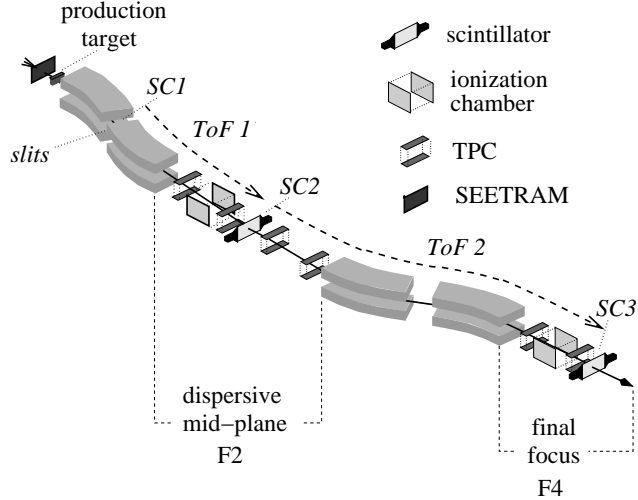,width=0.6\textwidth}
    \caption{\em A schematic view of the FRagment Separator (FRS) with the
      detection set-up. The isotope identification was possible in
      both sections of the spectrometer
      by measuring the time of flight (TOF) between scintillators (SC) and 
      the energy deposition in ionization chambers. Several
      position-sensitive detectors (TPC) allowed  tracking of projectiles and
      fragments and  momentum measurements of the fragments.}
\label{f:frs}
  \end{center}
\end{figure}

\begin{figure}
  \begin{center}
   \epsfig{file=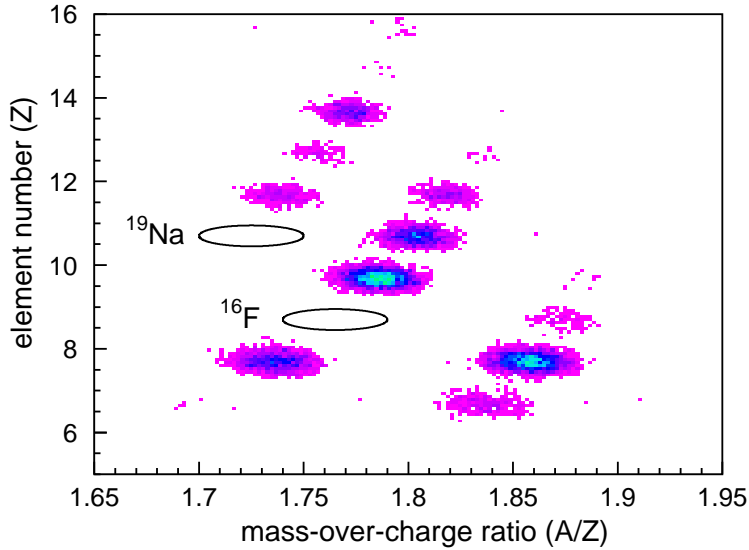,width=0.8\textwidth}
    \caption{\em An example of the fragment identification in the first half
      of the FRS. The isotope identification is confirmed by the non-observation of the unbound nuclei
      $^{19}${\rm Na} and $^{16}${\rm F}.}
\label{f:id}
  \end{center}
\end{figure}

Fragments were identified with respect to their nuclear charge, $Z$, and their mass-over-charge
ratio, $A/Z$, by a combination of energy-deposition, magnetic rigidity and time-of-flight (ToF) measurements.
The set-up used is shown in Fig.\ref{f:frs}. While in principle both observables can be measured
in both sections of the FRS, only the first half up to F2 was used to identify and count the
fragments.
To this end, the energy deposition in an ionization chamber at F2 was measured together
with the time of flight (ToF1) between scintillators ``SC1'' and ``SC2''. Positions and
angles at F2 could be determined with the help of two position-sensitive detectors (TPC).
The ToF was calibrated with the primary $^{36}$Ar beam from SIS with several
known energies which allowed to determine the length of the flight path and the offset.
Together with magnetic rigidity ($B\rho$) determined from Hall-probe measurements and corrections
for deviations from the optical axis, the ToF gave $A/Z$ of the fragments.
At the same time, the velocity-dependent energy deposition of the beam was recorded
and calibrated in terms of nuclear charge, $Z$.
Typical results of the isotope identification obtained are visualized in Fig.\ref{f:id}.
The non-observation of the unbound nuclei $^{19}$Na and $^{16}$F allowed to verify independently
the isotope identification.

\section{Data analysis and results}
\begin{figure}
  \begin{center}
   \epsfig{file=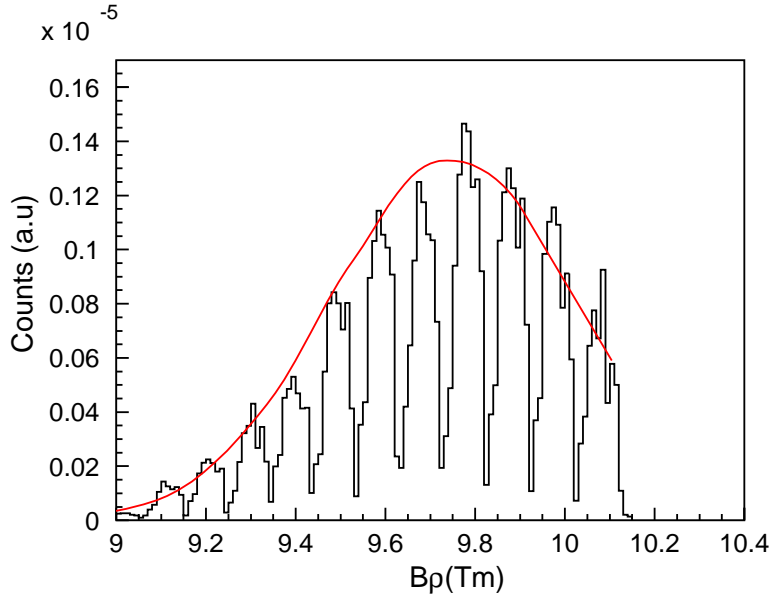,width=0.8\textwidth}
    \caption{\em Magnetic-rigidity distribution of the fragment $^{14}${\rm O}
      reconstructed from several normalized $B\rho$ settings. The envelope of the distribution
      is fitted to a Gaussian.}
\label{f:Br}
  \end{center}
\end{figure}

In the present experiment, the requirements of the main secondary-reaction study dictated
an experimental setup that was not optimum for the present purpose. The main limitation was the width of
the F1 start scintillator (``SC1'' in Fig.\ref{f:frs}) of only 2 cm in x- (bending) direction.
This leads to important cuts in the fragment longitudinal momentum distributions such that several
$B\rho$ settings had to be analyzed to cover the entire phase space distribution of each fragment.
Each $B\rho$ setting allowed to produce a two-dimensional spectrum as the one shown in
Fig.\ref{f:id}. For each isotope selected by a two-dimensional condition in Fig.\ref{f:id},
its x-distribution in the F2 momentum-dispersive focal plane was converted to a longitudinal
momentum distribution and properly normalized with the help of the SEETRAM beam current detector.
The reconstructed magnetic rigidity distribution of the isotope $^{14}$O is shown in Fig.\ref{f:Br}.

The reconstructed Gaussians for each isotope yield directly the average longitudinal momentum
in the laboratory frame and the momentum width. From their areas, production cross sections can be determined once the necessary corrections have 
been applied.

\subsection{Production cross-sections}

An important ingredient in determining the production cross sections is the transmission calculation.
In bending direction, a $B\rho$ scan allows to cover the entire longitudinal momentum distribution
of each fragment or at least to cover enough momentum space that a Gaussian fit to the data
is possible. The same is not possible, however, perpendicular to the bending plane. This means
that the angular acceptance for each isotope has to be calculated numerically, 
based on its measured longitudinal
momentum width and assuming that longitudinal and transverse momentum widths
are identical. The numerical procedures have been published in Ref.~\cite{Ben02}.
It is well known that the momentum widths depend approximately on the square root of the
mass difference between projectile and fragment (e.g. Ref.~\cite{morrissey}).
Consequently, the transmission of fragments close to the $^{36}$Ar projectile
(like the Ar or Cl fragments in Table~\ref{table1}) are transmitted with about 100\%, whereas the
isotopes $^{12,13}$N are transmitted only with about 33\% efficiency. 

\begin{figure}
  \begin{center}
   \epsfig{file=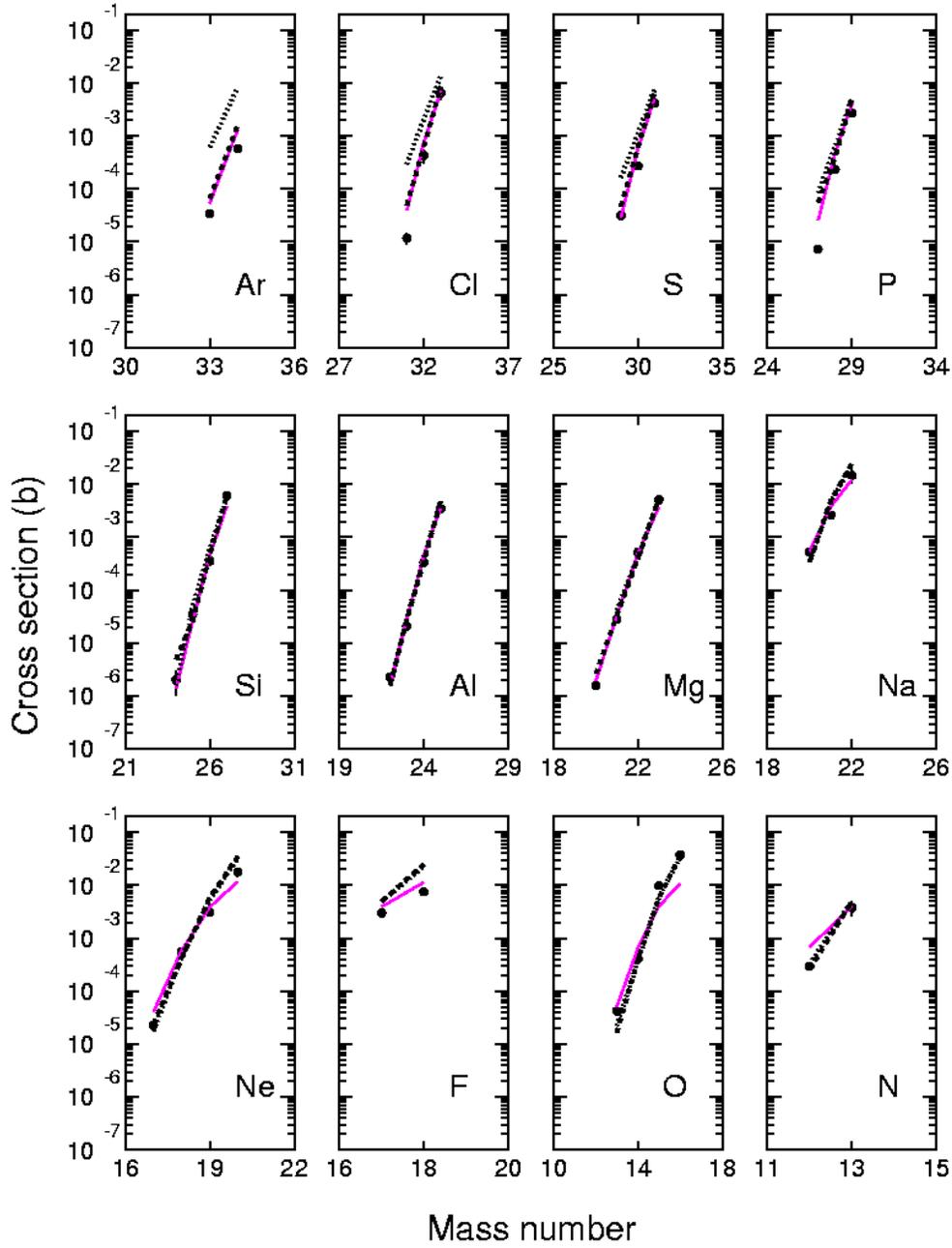,width=\textwidth}
    \caption{\em Production cross sections of neutron-deficient $^{36}$Ar fragments
      measured at 1050 $A$ MeV. Solid lines link {\sc EPAX}~{\protect\cite{epax}} predictions.
      Dotted and dashed lines depict model calculations with the {\sc ISABEL} intranuclear-
      cascade~{\protect\cite{Isabel}} and the 
      {\sc ABRABLA} abrasion-ablation model~{\protect\cite{abra}}, respectively.}
\label{f:Xsec}
  \end{center}
\end{figure}

Further corrections to the isotope yields are due to the dead time of the data acquisition
(approximately between 50\% and 70\%) and secondary reactions of the produced residues in the target (of about 10\%). Finally, the observed number of counts of each isotope
is converted to a cross section using the number of target atoms and the number of incident ions  
as determined from the SEETRAM beam intensity monitor. The resulting production cross sections
for 38 isotopes are listed in Table~\ref{table1} and visualized in Fig.\ref{f:Xsec}.

\unitlength0.7\textwidth 
\begin{table}[t]
\begin{center}
\caption{\em Production cross sections of neutron-deficient fragments from $^{36}${\rm Ar} 
      measured at 1050 $A$ MeV}
\protect\label{table1} 
\vspace*{0.5cm}
\begin{tabular}{|c|c|c|c|}
	\hline
	Isotope & $\sigma_{prod}$~(b)&Isotope & $\sigma_{prod}$~(b) \\\hline \hline
	$^{12}$N & $(2.97\pm 0.09)\cdot 10^{-4}$&$^{22}$Al & $(2.30\pm 0.30)\cdot 10^{-6}$ \\\hline 
	$^{13}$N & $(3.72\pm 1.19)\cdot 10^{-3} $& $^{23}$Al & $(2.10\pm 0.30)\cdot 10^{-5}$ \\\hline
	$^{13}$O & $(4.20\pm 0.20)\cdot 10^{-5}$ &$^{24}$Al & $(3.34\pm 0.13)\cdot 10^{-4}$ \\\hline 
	$^{14}$O & $(4.12\pm 0.13)\cdot 10^{-4}$ & $^{25}$Al & $(3.48\pm 0.52)\cdot 10^{-3}$ \\\hline
	$^{15}$O & $(9.82\pm 0.29)\cdot 10^{-3}$ & $^{24}$Si & $(2.00\pm 1.00)\cdot 10^{-6}$  \\\hline
	$^{16}$O & $(3.76\pm 0.64)\cdot 10^{-2}$ & $^{25}$Si & $(3.40\pm 0.10)\cdot 10^{-5}$ \\\hline
	$^{17}$F & $(2.99\pm 0.09)\cdot 10^{-3}$ &$^{26}$Si & $(3.52\pm 0.11)\cdot 10^{-4}$ \\\hline 
	$^{18}$F & $(7.58\pm 1.35)\cdot 10^{-3}$& $^{27}$Si & $(6.13\pm 1.04)\cdot 10^{-3}$\\\hline
	$^{17}$Ne & $(2.30\pm 0.10)\cdot 10^{-5}$ & $^{27}$P & $(7.40\pm 1.40)\cdot 10^{-6}$  \\\hline
	$^{18}$Ne & $(5.67\pm 0.17)\cdot 10^{-4}$ &$^{28}$P & $(2.33\pm 0.09)\cdot 10^{-4}$ \\\hline 
	$^{19}$Ne & $(3.12\pm 0.09)\cdot 10^{-3}$ &$^{29}$P & $(2.78\pm 0.17)\cdot 10^{-3}$ \\\hline 
	$^{20}$Ne & $(1.80\pm 0.40)\cdot 10^{-2}$&$^{29}$S & $(3.20\pm 0.10)\cdot 10^{-5}$ \\\hline 
	$^{20}$Na & $(5.24\pm 0.15)\cdot 10^{-4}$ &$^{30}$S & $(2.73\pm 0.11)\cdot 10^{-4}$ \\\hline
	$^{21}$Na & $(2.63\pm 0.38)\cdot 10^{-3}$ &$^{31}$S & $(4.17\pm 0.39))\cdot 10^{-3}$\\\hline 
	$^{22}$Na & $(1.47\pm 0.46)\cdot 10^{-2}$ & $^{31}$Cl & $(1.20\pm 0.30)\cdot 10^{-5}$ \\\hline 
	$^{20}$Mg & $(1.60\pm 0.30)\cdot 10^{-6}$&$^{32}$Cl & $(4.36\pm 1.32)\cdot 10^{-4}$ \\\hline  
	$^{21}$Mg & $(2.80\pm 0.50)\cdot 10^{-5}$ & $^{33}$Cl & $(6.59\pm 1.87)\cdot 10^{-3}$\\\hline  
	$^{22}$Mg & $(5.19\pm 0.21)\cdot 10^{-4}$& $^{33}$Ar & $(3.40\pm 0.10)\cdot 10^{-5}$\\\hline 
        $^{23}$Mg & $(5.13\pm 0.15)\cdot 10^{-3}$&$^{34}$Ar & $(5.73\pm 0.19)\cdot 10^{-4}$ \\\hline

\end{tabular}

\end{center}
\end{table}

The errors involved in the cross section determination are partly constant (2\% each for
the angular acceptance determination~\cite{Ben02} and beam intensity determination) and partly vary appreciably from
isotope to isotope. The main contribution comes from the errors of the Gaussian fit to
the reconstructed momentum distributions.
The total combined errors of the production cross sections vary between 3\% and
about 30\% with the exception of $^{24}$Si where the error is of about 50\%.
 
\subsection{Momentum distributions}
As mentioned earlier, the momentum distributions for each isotope result directly from
Gaussian fits to the reconstructed $B\rho$ distributions such as the one shown in Fig.~\ref{f:Br}.
The resulting laboratory momenta are then transformed to the projectile system.
As it is well known, the centroid values are small (on the order of 100-200 MeV/c),
i.e. the fragments have practically the same velocity as the projectile beam.
The central momenta decrease approximately linearly with mass loss from the projectile,
with a slope similar to what can be found in the Morrissey systematics~\cite{morrissey}.
Since the error bars in our measurements are rather large, and do not give more insight
than what can be found in the literature, we do not discuss this observable here.

\begin{figure}
  \begin{center}
   \epsfig{file=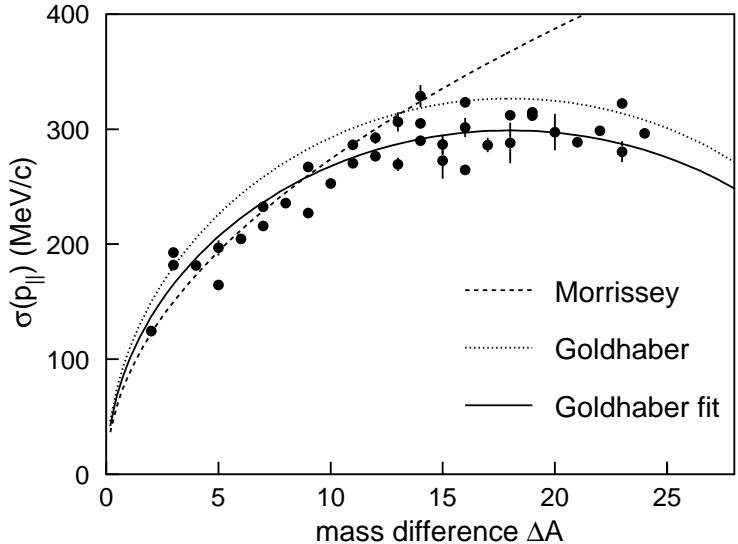,width=0.8\textwidth}
    \caption{\em Widths of the longitudinal momentum distributions plotted as a
      function of mass loss, $\Delta A$, from the projectile.
      The dashed curve shows the $\sqrt{\Delta A}$ dependence of the Morrissey systematics~\cite{morrissey}.
      The dotted curve follows Goldhaber's model~{\protect\cite{goldhaber}}.
 The full curve has been obtained by fitting the constant $\sigma_0$ in Goldhaber's model to the present data (see text).}
\label{f:Pwidth}
  \end{center}
\end{figure}

The widths of the longitudinal momentum distributions, $\sigma(p_{||}$),
are listed in Table~\ref{table2} and are shown graphically in
Fig.~\ref{f:Pwidth}. A marked deviation from the parabolic shape predicted by the
Morrissey systematics~\cite{morrissey} is clearly visible above a mass difference
of 10 units between fragment and projectile. 
\unitlength0.7\textwidth 
\begin{table}[t]
\begin{center}
\caption{\em Longitudinal momentum  widths in the projectile system of neutron-deficient 
fragments from $^{36}$Ar measured at 1050 $A$ MeV  }
\protect\label{table2} 
\vspace*{0.5cm}
\begin{tabular}{|c|c|c|c|}
	\hline
	Isotope & $\sigma(p_{||})$~(MeV/c)& Isotope & $\sigma(p_{||})$~(MeV/c)\\\hline \hline
	$^{12}$N & $296\pm 2$ & $^{23}$Al & $269\pm 6$\\\hline
	$^{13}$N &  $322\pm 3$& $^{23}$Mg & $306\pm 8$\\\hline
	$^{13}$O &  $281\pm 10$& $^{24}$Al & $293\pm 5$\\\hline
	$^{14}$O &  $299\pm 1$&$^{24}$Si &  $277\pm 4$\\\hline
	$^{15}$O & $289\pm 4$&$^{25}$Al & $286\pm 2$\\\hline
	$^{16}$O & $297\pm 16$&$^{25}$Si &  $270\pm 3$\\\hline
	$^{17}$F &  $315\pm 4$&$^{26}$Si &  $253\pm 2$\\\hline
	$^{17}$Ne &  $312\pm 3$&$^{27}$P &  $227\pm 4$\\\hline
	$^{18}$F &  $288\pm 18$&$^{27}$Si & $267\pm 3$\\\hline
	$^{18}$Ne &  $312\pm 3$&$^{28}$P & $236\pm 3$\\\hline
	$^{19}$Ne &  $286\pm 6$&$^{29}$P & $216\pm 2$\\\hline
	$^{20}$Mg &  $265\pm 4$&$^{29}$S & $233\pm 2$\\\hline
	$^{20}$Na &  $323\pm 4$&$^{30}$S & $205\pm 2$\\\hline
	$^{20}$Ne &  $302\pm 8$&$^{31}$Cl & $165\pm 2$\\\hline
	$^{21}$Mg &  $287\pm 8$&$^{31}$S & $197\pm 7$\\\hline
	$^{21}$Na &  $273\pm 16$&$^{32}$Cl & $182\pm 3$\\\hline
	$^{22}$Al &  $290\pm 1$&$^{33}$Ar & $182\pm 4$\\\hline 
	$^{22}$Mg &  $305\pm 4$&$^{33}$Cl & $193\pm 2$\\\hline
	$^{22}$Na &  $329\pm 10$&$^{34}$Ar & $124\pm 3$\\\hline	

\end{tabular}

\end{center}
\end{table}

\section {Discussion}
\subsection{Production cross- sections}

The present data constitute an excellent data base to benchmark the
predictive power of different fragmentation models at the limit of unbound
nuclei. At these energies, the fragmentation process is described in terms of
a two-step model. In the first step, the collision between the nucleons in
the overlap region between projectile and target leads to the formation of
two residues of both reaction partners called prefragments, with similar
kinematical properties, smaller mass and a given excitation energy. In the
second step, the residues of both reaction partners release their energy
excess by evaporating nucleons and $\gamma$-rays. The proximity of the investigated
nuclei to the proton drip line is expected to increase the sensitivity of
their production cross sections to the excitation energy induced in the first
stage of the collision or the competition between proton and neutron
evaporation during the second stage.

To investigate the reaction mechanism leading to the formation of the
measured nuclei we have performed calculations using two different models to
describe the first stage of the reaction, the intranuclear-cascade model
{\sc ISABEL} \cite{Isabel} and the abrasion model of Gaimard and Schmidt \cite{abra}. Both models were
coupled to the {\sc ABLA} evaporation code \cite{Jun98} to describe the second stage of the
reaction. These two models follow completely different approaches. In the
intranuclear-cascade model {\sc ISABEL}, the interaction between projectile and
target is described as a sequence of elastic or inelastic nucleon-nucleon
collisions. The moving nucleons follow classical trajectories and the only
quantum mechanical ingredient is the Pauli blocking.
Nucleons with sufficiently high momenta can leave their respective nuclear volumes,
the leftover nucleons form the prefragment. 
The final prefragment excitation energy is obtained from particle-hole 
excitations in the initial Fermi distribution of the removed nucleons 
plus the energies of the scattered
nucleons that end up below the ``cutoff energy''; the latter has been introduced to
terminate the history of cascade nucleons with energies too low to escape their
respective potential wells~\cite{Isabel}.

The abrasion model follows geometrical considerations, the size of the
projectile and target prefragments is given by the non-overlapping region
between both partners, which is defined by the impact parameter. The
excitation energy induced in the collision is determined by the energy of the
holes of the removed nucleons, assuming a Fermi distribution of nucleons
inside the nucleus, which has been empirically demonstrated to be on average
around 27 MeV per abraded nucleon~\cite{excita}.

Other than the physical models discussed above, the empirical parametrization 
of fragmentation cross sections, {\sc EPAX}~\cite{epax}, has little physical 
input and merely aims at reproducing measured data by numerical expressions. 
Such an approach is nevertheless useful e.g. for planning experiments or for 
cases where fast iterative calculations of secondary cross sections are 
needed. In the present context, the {\sc EPAX} formula is particularly useful 
since it contains a parametrization of the ``memory effect'', i.e. the 
influence of the proton excess of the projectile on the proton excess of the 
fragments. As explained in detail in Ref.~\cite{epax}, the numerical form of 
this ``memory effect'' has been adjusted to measured $^{58}$Ni fragmentation 
cross sections. A comparison with the present high-precision $^{36}$Ar cross 
sections can serve as a check of the predictive power of {\sc EPAX} for 
proton-rich projectiles.

In figure~\ref{f:Xsec} we compare the measured production cross sections with the
results of the calculations obtained with the intranuclear-cascade model
{\sc ISABEL} (dotted lines) and the abrasion model of Gaimard and Schmidt {\sc ABRA} 
(dashed lines) coupled with the {\sc ABLA} evaporation model. 
The predictions obtained with the semi-empirical parametrization {\sc EPAX} are included in this figure as well (solid line) .

In general the three calculations provide a very good description of the
measured cross sections even for weakly bound nuclei at the proton
drip line. Only a detailed comparison shows some trends. The {\sc ISABEL}
calculation seems to overestimate the production cross sections of
heavy-projectile residues while the abrasion model provides an excellent
description of all the data. Since both calculations use the same
evaporation model the observed deviations must be due to the
different models used to describe the first stage of the collision. In
addition, residues close to the projectile do not reach the limiting
fragmentation regime and consequently their production cross section is
sensitive to the entrance channel. The observed overestimation
of the production cross section with {\sc ISABEL} can be explained if one 
considers that this code produces fragments with 20\% less excitation energy on average than the {\sc ABRA} code. 
As pointed out, the production of heavy projectile residues is very much affected by the first stage of the
collision. In particular, a lower excitation energy leads to shorter
evaporation chains and consequently an increase of the production cross
section of heavy residues. The good description of these data with the abrasion-ablation model ABRABLA confirms its predictive power and validates the physical assumptions used in this code.\\ 

The overall agreement of the measured
production cross sections with the EPAX formula~\cite{epax} is excellent, 
as can be seen in Fig.\ref{f:Xsec}.
The present experiment reaches the proton drip line up to $Z=13$, Al. Even
drip-line nuclei are produced with cross sections that lie well on the EPAX predictions.
This confirms observations that were made for another
neutron-deficient projectile, $^{58}$Ni~\cite{blank}. 

\begin{figure}
  \begin{center}
   \epsfig{file=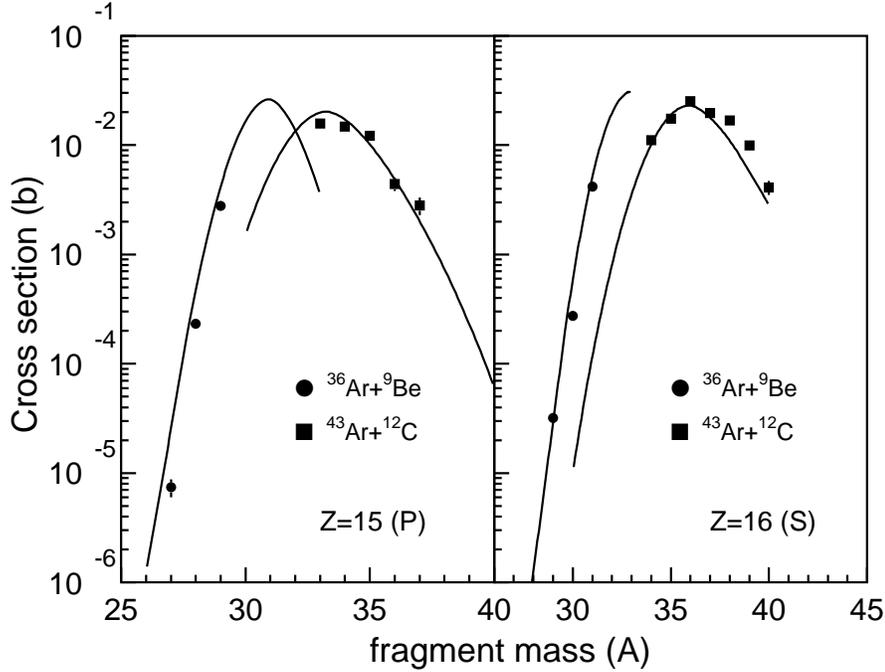,width=0.95\textwidth}
    \caption{\em Production cross sections of P isotopes (left) and S isotopes (right)
      from the fragmentation of $^{36}${\rm Ar}+$^{9}${\rm Be} at 1050 $A$ MeV (circles, this work)
      in comparison to data from $^{43}${\rm Ar}+$^{12}${\rm C} fragmentation at 222 $A$ MeV
      (squares, Ref.~{\protect\cite{Wan99}}). The curves denote the respective {\sc EPAX} predictions.}
\label{f:ar3643}
  \end{center}
\end{figure}

Our new data for $^{36}$Ar ($N/Z=1.0$) allow to test the EPAX predictions for the production of exotic nuclei
over a wide range of projectile-$N/Z$ values by comparing them to similar data from $^{43}$Ar ($N/Z=2.39$)
on $^{12}$C at 222 $A$ MeV~\cite{Wan99}. We assume that for both incident energies, 222 and 1050 $A$MeV,
the cross sections are largely energy-independent, so that EPAX can be applied. Fig.~\ref{f:ar3643} demonstrates that
at least the proton-rich slopes of the $^{36}$Ar-induced distributions
and the neutron-rich slopes of the $^{43}$Ar-induced distributions are correctly
predicted by EPAX. Similarly good agreement was found for isotope distributions from
$^{96}$Ru and $^{96}$Zr fragmentation~\cite{rnb5}. 

\subsection{Momentum distributions}
A precise knowledge of the kinematics of high-energy projectile fragmentation is essential to correctly predict
the yields of secondary beams from an in-flight separator like the FRS.
We have discussed in subsect.{\it 3.1} how the transmission of the FRS is affected by the
momentum distributions.

In the traditional two-step model of high-energy projectile fragmentation, the momentum widths are
largely determined by the random addition of the Fermi momenta of the nucleons
that are knocked out in the first (``abrasion'') phase of the reaction.
The corresponding prefragment momentum widths have been given by Goldhaber~\cite{goldhaber} as
\begin{equation}
\sigma(p_{||})=\sigma_0 \sqrt{\frac{A_{pf} ( A_p - A_{pf} )}{A_p - 1}}
              =\frac{p_F}{\sqrt{5}} \sqrt{\frac{A_{pf} ( A_p - A_{pf} )}{A_p - 1}}
\label{eq:gold}
\end{equation}
where $A_p (A_{pf})$ refers to the projectile (prefragment) mass; the latter 
cannot be observed. The quantity $p_F$ denotes the Fermi momentum of the 
nucleons in the projectile. Subsequent evaporation (``ablation'') of nucleons 
in the second phase additionally broadens the momentum distributions somewhat,
but the most important effect here is that the number of nucleons is reduced, 
so that the original ``Goldhaber'' distribution becomes flatter and can be 
better described (at least for fragments close to the projectile mass) by 
Morrissey's formula~\cite{morrissey}:
\begin{equation}
\sigma(p_{||})=\frac{150} {\sqrt{3}} \sqrt{A_p - A_f }
\label{eq:morrissey}
\end{equation}
where now $A_f$ denotes the (observed) fragment mass.
Systematic studies of the validity of the Morrissey systematics 
have been undertaken about a decade ago at the FRS (e.g. Ref.~\cite{Han93}).

Up to about a mass loss of 10 units, the present data follow nicely the 
Morrissey systematics (dashed curve in Fig.~\ref{f:Pwidth}).
For fragments further removed from the projectile, however, the widths tend to 
saturate. Similar observations were made in a comprehensive study of $^{86}$Kr 
fragmentation at 500 $A$ MeV~\cite{weber}, where it was found that the widths 
even became smaller again for fragments with masses smaller than about half 
the projectile mass. One practical solution is to keep the Goldhaber formula 
(Eq.~\ref{eq:gold}) but fit the width coefficient $\sigma_0$
to the measured data of Fig.~\ref{f:ar3643}
(full line). This leads to a parameter of $\sigma_0^f = 98.2 \pm 0.2$~MeV/c, 
smaller than the Goldhaber prediction (for {\em prefragments})
of $\sigma_0^{pf} = 112$ MeV/c. The latter value is based on a Fermi momentum 
in $^{36}$Ar of $p_F =~$250 MeV/c~\cite{moniz}. The smaller $\sigma_0^f$ value 
reflects the evaporation mass loss as discussed above. It compares well with 
other data found in the literature for light nuclei, e.g. 
Viyogi {\it et al.}~(Ref.~\cite{viyogi}) found $\sigma_0 = 94 \pm 5$~MeV/c 
for $^{40}$Ar at 213 $A$MeV.
  
\section{Summary}

We have studied the fragmentation of neutron-deficient $^{36}$Ar at 
relativistic energies and have obtained precise values for production cross 
sections and longitudinal momentum widths of very neutron-deficient fragments, 
reaching the proton-drip line up to aluminum. Both observables can be 
reproduced well by their respective empirical parametrizations: the cross 
sections match excellently those predicted by the {\sc EPAX} formula, thus 
confirming its validity also for the case of very neutron-deficient 
projectiles. The longitudinal momentum widths are best fitted by a modified 
Goldhaber formula, where the width constant $\sigma_0$ is replaced by an 
empirical value. The measured production cross sections are also compared with 
physical-model calculations performed with (i) the {\sc ISABEL} 
intranuclear-cascade model, and (ii) the Gaimard-Schmidt abrasion model, both 
coupled to a standard evaporation code ({\sc ABLA}). The {\sc ISABEL} code 
gives an overall good description of the data however, some discrepancies are 
observed for the heavier residues. 
For these nuclide very close to the projectile the {\sc ISABEL} code clearly 
overestimate the production cross sections. This result is interpreted as a 
consequence of an underestimation of the excitation energy induced in the 
reaction; lower excitation energies lead to shorter evaporation chains 
which increase the production of residues close to the projectile. The 
abrasion model {\sc ABRABLA} provide an extremely good description of all the data.

These results indicate that in the range of fragment masses studied here the 
assumption of a two-stage model works well; the model consists of a fast 
removal of quasi-free nucleons during the geometrical overlap of projectile 
and target, followed by evaporation of particles from a thermally 
equilibrated prefragment. In addition, we confirm the predictive power of the 
abrasion code that can also be used to describe the production of residual 
nuclide in more complex reactions where the EPAX formula can not be applied, as it is the case for fissile projectiles.

\section{Acknowledgements}

The authors wish to express their gratitude to their colleagues from the accompanying
experiments for their help with the experimental setup.
They also acknowledge technical support from A.~Br{\"u}nle, K.H.~Behr, and K.~Burkard. 
This work was supported by the EU under contract ERBFMGECT95
0083, and by the Spanish Ministery of Science and Technology  
under contract number FPA2001-0144-C05-4-01.

\end{document}